# Tailoring stresses in piezoresistive microcantilevers for enhanced surface stress sensing: insights from topology optimization

Chao Zhuang[1,2,*], Kosuke Minami[2], Kota Shiba[2], and Genki Yoshikawa[1,2,*]

[1]*Materials Science and Engineering, Graduate School of Pure and Applied Science, University of Tsukuba, 1-1-1 Tennodai, Tsukuba, Ibaraki, 305-8571 Japan*
[2]*Research Center for Macromolecules and Biomaterials (RCMB), National Institute for Materials Science (NIMS), 1-1 Namiki, Tsukuba Ibaraki 305-0044 Japan*

E-mail: zhuang.chao@nims.go.jp; yoshikawa.genki@nims.go.jp.

In assessing piezoresistive microcantilever sensitivity for surface stress sensing, the key is its capacity to translate surface stress into changes in resistance. This change hinges on the interplay between stresses and piezoresistivity. Traditional optimization has been constrained by rudimentary 1D models, overlooking potentially superior designs. Addressing this, we employed topology optimization to optimize Si(100) microcantilevers with a p-type piezoresistor. This led to optimized designs with up to 30% enhanced sensitivity over conventional designs. A recurrent "double-cantilever" configuration emerged, which optimizes longitudinal stress and reduces transverse stress at the piezoresistor, resulting in enhanced sensitivity. We developed a simplified model to analyze stress distributions in these designs. By adjusting geometrical features in this model, we identified ideal parameter combinations for optimal stress distribution. Contrary to conventional designs favoring short cantilevers, our findings redefine efficient surface stress sensing, paving the way for innovative sensor designs beyond the conventional rectangular cantilevers.

## 1. Introduction

Piezoresistive microcantilevers are a promising sensing platform that has been applied across disciplines because of their compactness and real-time operation capability.[1–3] Upon chemical or physical interactions with target stimuli, surface stress is generated at the cantilever surface. The surface stress bends the microcantilever and leads to a change in resistance at the embedded piezoresistor.[4] These devices are particularly advantageous as they can operate in opaque medium and are thus suitable for a wide range of sensing applications, such as bioassays[5] and environment monitoring.[6] However, piezoresistive readout suffers from a low signal-to-noise ratio because of the low sensitivity and the high electronic noise. Although previous studies have suggested several strategies to reduce electronic noise by miniaturization[7] and optimizing boron doping,[8] low sensitivity has been the major bottleneck to the broader applications of this technology.

The sensing signal of a piezoresistive microcantilever is given by the product of the stress tensor and the piezoresistive tensor at the piezoresistor under unit surface stress loading.[9] Since the piezoresistance coefficients, can be either positive or negative depending on the material,[10] maximizing the sensitivity thus requires stress maximization in some directions while minimization in others. The complexity involved has made geometry optimization that takes all stress components into account a formidable task. Traditional optimization has relied on simple one-dimensional models, such as Stoney's equation[11] and Timoshenko's bimetal strip.[12] These studies suggest a general rule of stiffness minimization, but they provide limited insights for piezoresistive readout, where consideration of stresses in different directions is required. Two-dimensional models based on the plate theory have provided approximate solutions for the deflection and stress distribution for cantilevers with simple geometries,[13–15] but they do not allow further generalization to complex geometries. Therefore, finite element analysis (FEA) has been used to study the complex relationship between stress and geometry. A series of studies have found that in contrast to the design of force-sensing cantilevers where a high-aspect-ratio design is preferred, surface stress sensing in general prefers a low-aspect-ratio geometry.[16] Moreover, topological features such as stress-concentrating holes[17] are pointed out to be effective in improving sensitivity. However, these numerical studies are far from systematic because they consider only cantilevers with primitive shapes, which imposes a strong constraint on the possible stress configurations and overlooks geometric features that can potentially generate superior sensitivity.

To obtain the optimal stress distribution and fully explore the design space of piezoresistive microcantilevers, a systematic optimization approach is needed. Density-based topology optimization is a powerful design method that has been applied in solving optimization problems in structural mechanics and specifically, the design of microelectromechanical systems (MEMS), such as nanomechanical resonators,[18] compliant mechanisms,[19] Radio Frequency MEMS switches,[20] and piezoresistive force sensors.[21] Topology optimization had also been applied by Pederson[22] to explore the optimal cantilever geometry for surface stress sensing, but the optimized designs obtained at the time suffered from poor manufacturability because of the one-node-connection problem, which is a common numerical artifact in displacement/stress maximization problems.[23]

In this study, we aim to explore highly sensitive while manufacturable designs of piezoresistive microcantilevers by using topology optimization integrated with the robust formulation. The robust formulation transforms the topology optimization into a minimax problem where the algorithm simultaneously optimizes for the sensitivity of eroded, intermediate, and dilated designs such that the final design is robust to fabrication error.[24] This technique has been applied and found to be effective in multiple structural mechanics optimization problems in eliminating one-node connections.[25] Focusing on microcantilevers with piezoresistors made from p-type silicon oriented in [110] direction, we successfully generate a series of optimized designs for various shapes of piezoresistors, outperforming rectangular benchmark designs. Importantly, since the optimized geometries generated by topology optimization are complex, we perform a systematic investigation on these geometries and arrive at an effective design guideline for maximal surface stress sensitivity, that is, a "double-cantilever" configuration, where the piezoresistor locates at the center of two connected cantilevers. This model provides a unique solution to the surface stress sensing problem, allowing for more efficient geometry design in future sensor development.

## 2. Numerical Methods

*2.1 Topology Optimization*

Density-based topology optimization is commonly structured within a designated domain, denoted as $\Omega$, accompanied by specific boundary and loading conditions. The optimization process involves the iterative adjustment of material density to optimize one or more objective functions $f$.[26] The domain is discretized into finite elements, and the presence or absence of materials is denoted by a density vector for each element, which also serves as the optimization variable. Element density values range between 0 (absence of material) and 1 (presence of material). The material attributes and biaxial stress are influenced by the element density and are exponentiated by a penalization factor, $p$, based on the Simple Isotropic Material with Penalization (SIMP) method.[27] This ensures designs with densities of only 0 or 1. Typically, p is set to 3, contingent on the Poisson's ratio of the material.[28]

The optimization involves: (1) Evaluate the objective function using FEA. (2) Determining the sensitivity of objectives and constraints concerning each element density variation using FEA and related adjoint solutions. (3) Adjusting the densities of all elements based on the sensitivity analysis from step (2), utilizing the Method of Moving Asymptotes (MMA). (4) Repeating the above steps until convergence criteria are satisfied or a set iteration limit is reached.[29]

*2.1.1. Filters and Projections*

To counteract the known numerical instabilities of checkerboarding[31] and mesh dependency[32], a Helmholtz filter[30] is employed. This involves computing the elemental density as an average, weighted by neighboring elements within a specific radius, $r_{min}$, during each iteration:

$$\tilde{\rho}_j - r_{min}^2 \left( \frac{\partial^2 \tilde{\rho}_j}{\partial x^2} + \frac{\partial^2 \tilde{\rho}_j}{\partial y^2} \right) = \rho_j, \qquad (1)$$

where $\rho_j$ and $\tilde{\rho}_j$ are the original and filtered density of element $j$, respectively.

The density filter introduces zones transitioning between solid and void areas, which lack physical relevance. Therefore, a Heaviside projection function is applied to remove intermediate densities.[33] The projected density is given by:[34]

$$\bar{\tilde{\rho}}_j = \frac{\tanh(\beta\eta) + \tanh\left(\beta(\tilde{\rho}_j - \eta)\right)}{\tanh(\beta\eta) + \tanh(\beta(1 - \eta))}, \qquad (2)$$

where $\beta$ represents the projection slope, and $\eta$ represents the projection threshold, which determines the density at which the projection function applies.

*2.1.2. Robust Formulation*

For ensuring a minimum feature size in both solid and void areas, we introduce the robust formulation which evaluates the objective functions of designs that are dilated, intermediate, and eroded in each iteration simultaneously. corresponding to $\eta = \eta^* < 0.5$, $\eta = 0.5$, and $\eta = 1 - \eta^*$, respectively. The MMA solver then focuses on optimizing the least efficient design among these. By addressing this minimax problem, the optimized design is imposed with a minimum length scale, effectively removing one-node connections and enhancing the design resilience against fabrication errors.[23–25]

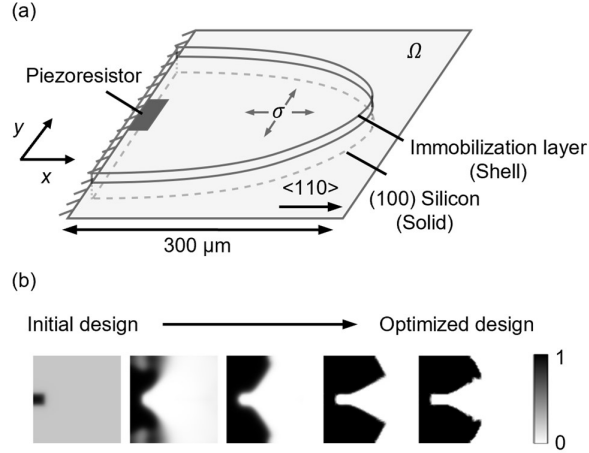

**Fig. 1.**

*2.1.3. Finite Element Model*

The cantilever bending is modeled as a bimorph consisting of an immobilization layer and a p-type (100) silicon layer with a fixed-free boundary condition shown in Fig. 1(a). The optimization algorithm iteratively modifies the material distribution to maximize the sensitivity. The design evolution is shown in Fig. 1(b). The immobilization layer is the functional layer that binds to probe molecules such as proteins and antibodies,[35] while the (100) silicon has been the conventional material used for sensor fabrication. The surface stress in the immobilization layer is modeled by a biaxial stress $\sigma_s$, which results in bending deformation that can be solved by:

$$Ku_\sigma = F_\sigma, \qquad (3)$$

where $F_\sigma$ denotes the force vector that contains the bending moment load induced by the surface stress, $K$ is the design-dependent stiffness matrix, and $u_\sigma$ is the deformation vector under surface stress. The constitutive relation yields the stress in the microcantilever:

$$\sigma = \frac{1}{2} C[(\nabla u_\sigma)^T + \nabla u_\sigma], \qquad (4)$$

where $C$ is the stiffness matrix of the material.[36]

Assuming plane stress condition, the resistance change of the piezoresistor with an area $A$ is given by the averaged tensor product between stress and piezoresistance. By further assuming the longitudinal and transverse directions of the piezoresistor aligns with the $x$ and $y$ direction of the design domain, we have the following expression for piezoresistor signal output:[37]

$$\overline{\Delta R/R} = \iint (\pi_l \sigma_{xx} + \pi_t \sigma_{yy}) dA. \qquad (5)$$

The subscripts denote tensor indices according to the Voigt matrix notation. $\pi_l$, $\pi_t$, $\sigma_{xx}$, and $\sigma_{yy}$ are the piezoresistance coefficients and stresses at the silicon surface along the longitudinal (*x*) and transverse (*y*) directions, respectively. Depending on the wafer orientation, the two piezoresistance coefficients are given by:

$$\pi_l = \pi_{11} - 2(\pi_{11} - \pi_{12} - \pi_{44}) \times (l_1^2 m_1^2 + m_1^2 n_1^2 + n_1^2 l_1^2) \quad (6)$$

and

$$\pi_t = \pi_{12} + (\pi_{11} - \pi_{12} - \pi_{44}) \times (l_1^2 l_2^2 + m_1^2 m_2^2 + n_1^2 n_2^2), \quad (7)$$

where $l$, $m$, and $n$ are components of the rotation matrix given by the Euler angles ($\phi,\theta,\psi$) associated with the wafer plane and orientation,[38] and $\pi_{11}$, $\pi_{12}$, and $\pi_{44}$ are piezoresistance coefficients of silicon. The same rotation matrix is used to rotate the stiffness tensor to account for the anisotropic elasticity of silicon at the [110] orientation.

2.1.4. Problem Formulation

The *p*-type piezoresistor is modeled near the clamped end and regarded as a passive area with a fixed density of 1. The material properties of the immobilization layer and silicon above and below the design domain are mapped through a general extrusion function from the design domain. Therefore, the presence of an element in the design domain represents simultaneously the presence of the immobilization layer and silicon with predefined thicknesses and local surface stress. The optimization objective is defined to be the sensitivity, that is, the resistance changes at the piezoresistor divided by unit surface stress $\sigma_s$:

$$f(\tilde{\bar{\rho}}) = \frac{\overline{\Delta R/R}}{|\sigma_s|}. \quad (8)$$

The optimization problem can be described by the following minimax problem:

$$\underset{\rho}{Max}: min\{f(\tilde{\bar{\rho}}^e(\rho)), f(\tilde{\bar{\rho}}^i(\rho)), f(\tilde{\bar{\rho}}^d(\rho))\}, \quad (9)$$

where the superscripts of $e$, $i$, and $d$ denote eroded, intermediate, and dilated designs. The volume constraint is applied to the dilated design.

2.1.5. Optimization Parameters

The topology optimization is implemented in the commercial FEA software COMSOL Multiphysics 6.1. The optimization is performed using the optimization module with MMA as the optimizer. The finite element analysis is performed using the structural mechanics module, in which a coupled interface of shell and solid are used to model the stressed immobilization layer and the silicon structural layer, respectively. The model is discretized by first-order quadrilateral elements with 50 elements a side in the in-plane directions. The filter size of $r_{min}$ is set to be 1.2 times the mesh size. Moreover, three elements are assigned in the thickness direction to prevent shear locking. This through-thickness mesh density is chosen because a further increase in density does not lead to significant changes in optimized designs. Only half of the geometry is modeled due to symmetry. A volume constraint of 0.6 is imposed on the dilated design to regularize the final design. This treatment is a common practice in solving optimization problems as it prevents the formation of floating materials that are not physically meaningful.[39] To ensure a sharp 1/0 design, a continuation method involves ramping up the projection slope $\beta$ from 1 to 16 by doubling in every 20 iterations is used.[20] The projection threshold $\eta$ is set to be 0.3, 0.5, and 0.7 for dilated, intermediate, and eroded designs, respectively. The final design is taken to be the intermediate design. After the optimization, the optimized design is reimported into a verification model with two times denser quadratic serendipity elements to calculate the accurate sensitivity of optimized designs.

*2.1.6. Finite Element Model Parameters*

The immobilization layer is assumed to be isotropic and its elastic properties are described by Young's modulus and Poisson's ratio listed in Table I along with other parameters used in the model. The *x* direction is oriented along the [110] direction in (100) silicon, which is associated with the Euler rotation angle of $(0,0,\pi/4)$ relative to the [100] direction where silicon exhibits cubic symmetry and its elasticity and piezoresistivity are described only three independent variables[38] as shown in Table II. In the [110] direction, the stiffness matrix becomes orthotropic and it is tabulated in a previous report.[40] The piezoresistance coefficients are calculated to be $\pi_l = 71.8 \times 10^{-11}$ Pa$^{-1}$, $\pi_t = -66.3 \times 10^{-11}$ Pa$^{-1}$ for *p*-type silicon used here.[10] According to Eq. (5), the different signs of piezoresistance coefficients indicate that the sensitivity will be maximized when the difference between stresses in the *x* and *y* directions is maximized.

After optimization, the optimized design will be verified using two times denser quadratic meshes to obtain accurate sensor performance.

**Table I.** Model and optimization constants

| Constants | Value | Unit |
|---|---|---|
| Silicon thickness | 3 | mm |
| Silicon Poisson's ratio | 0.27 | 1 |
| Surface stress, $\sigma_s$ | −1 | N·m$^{-1}$ |
| Immobilization layer thickness | 50 | nm |
| Immobilization layer Young's modulus | 80 | GPa |
| Immobilization layer Poisson's ratio | 0.42 | 1 |
| Piezoresistor area | 1600 | mm$^2$ |
| Design domain width | 300 | mm |
| Maximum mesh size | 6 | mm |
| Filter radius, $r_{min}$ | 7.2 | mm |

**Table II.** The stiffness and piezoresistance coefficients of (100) silicon in [100] orientation used for tensor rotation.

| Stiffness Coefficients (GPa)[a] | | Piezoresistance Coefficients ($10^{-11}$ Pa$^{-1}$)[b] | |
|---|---|---|---|
| $c_{11}$ | 165.7 | $\pi_{11}$ | 6.6 |
| $c_{12}$ | 63.9 | $\pi_{12}$ | −1.1 |
| $c_{44}$ | 79.56 | $\pi_{44}$ | 138.1 |

[a] The stiffness coefficients are obtained from reference.[41] [b] The piezoresistance coefficients are obtained from reference.[10]

*2.2. Benchmark Model*

To verify the effectiveness of topology optimization, rectangular microcantilevers with simply optimized geometries are used for benchmarking. The length of microcantilevers is longer than the piezoresistor by only one mesh to avoid high-stress regions near the edge. The width is set to be the same as the design domain. Such a design is used for comparison because previous studies have pointed out that cantilevers with short and wide geometries have a higher sensitivity to surface stress, in contrast to the long and thin geometries that are preferred in force sensing.[42] The benchmark models are evaluated with the same quadratic mesh density elements as the verification model right after the optimization. Mesh convergence analysis is performed to ensure the model has yielded the correct results.

## 3. Results and Discussion

As an illustration of topology optimization, the optimized design for a 40 μm × 40 μm piezoresistor and the benchmark model is shown in Fig. 2(a), where the square frame at the left edge represents the prescribed piezoresistor. The optimized design features a "double-cantilever" configuration where the piezoresistor locates at the junction between two combined cantilevers. Compared to the benchmark model, which has been shown in previous optimization studies to be optimal for rectangular designs,[42] the optimized design from topology optimization exhibits a concentrated piezoresistive signal profile at the piezoresistor as shown in Fig. 2(b). Since the piezoresistance coefficients along the *x* and *y* directions are of similar magnitude but with opposite signs, which in turn requires stresses along both directions to be maximally different to maximize the piezoresistive sensitivity as is suggested in Sect. 2.1.6 and Eq. (5). Therefore, the concentration of the piezoresistive signal is a result of the synergy of maximal $\sigma_{xx}$ and minimal $\sigma_{yy}$ at the piezoresistor as shown in Figs. 2(c)–(d). In contrast, such a synergy does not exist in the benchmark model due to the lack of design flexibility of straight edges. As a result, the benchmark model exhibits a lower sensitivity compared to designs obtained from topology optimization.

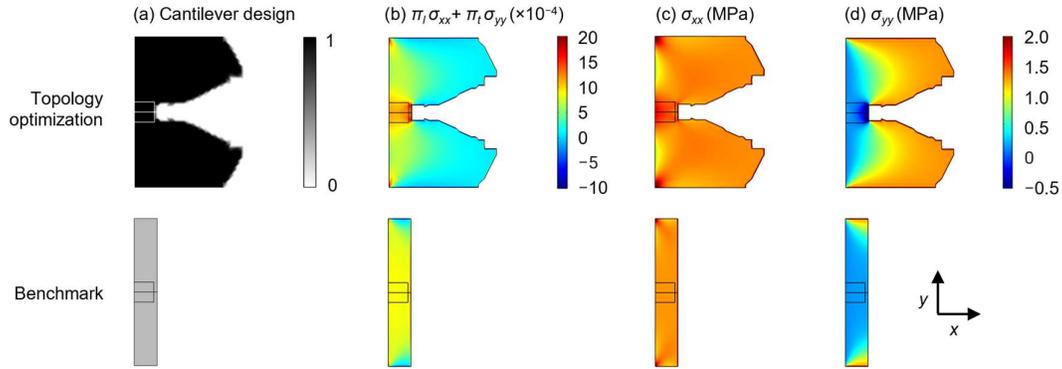

**Fig. 2.**

*3.1. The Effects of Piezoresistor Shape*

To investigate the effects of the piezoresistor shape on the optimized design, topology optimization is performed on piezoresistors with a series of aspect ratios (the ratio between length and width) while fixing the total area to be 1600 μm² (the area of the whole design domain being 300 μm × 300 μm). The relatively small area ensures that there are sufficient degrees of freedom for the optimization algorithm to reach the global optimum, otherwise, it will lead to suboptimal designs that have a sensitivity lower than the benchmark model (Figs. S1 and S2 in the supplementary materials). As the aspect ratio of the piezoresistor increases from 1/8 to 8, the corresponding optimized designs are shown in Fig. 3(a). When the aspect ratio is low, aside from the prominent feature of the double-cantilever, an additional small cantilever emerges out of the center of the piezoresistor, which creates two regions of stress concentration instead of one to compensate for the flat shape of the piezoresistor. As the aspect ratio increases, the small cantilever disappears, and all designs feature the double-cantilever with increasing shape regularity. Note that the irregular edges in the 1/4 and 1/2 designs do not contribute to sensor performance. Instead, they are a result of the flat optimization landscape at a high cantilever length. These irregularities can be eliminated by a smaller volume constraint that penalizes the extra materials that are irrelevant to sensor performance.

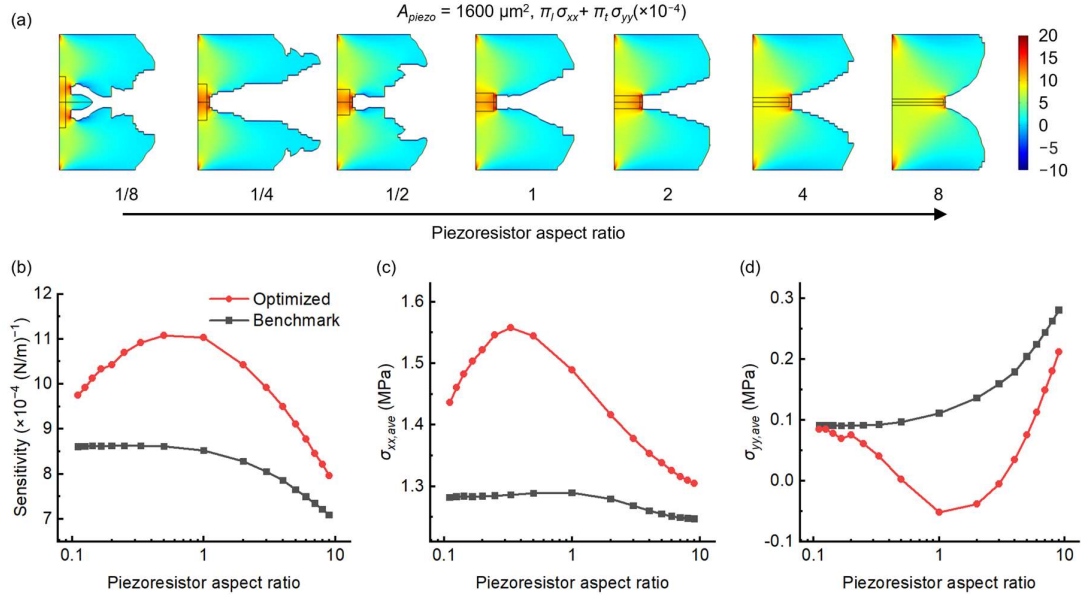

**Fig. 3.**

To evaluate the performance of optimized designs, we calculate their sensitivity by averaging the piezoresistive signal over the piezoresistor. The calculated sensitivity is compared with the benchmark model as shown in Fig. 3(b). The optimized designs in general outperform the benchmark across different piezoresistor aspect ratios. The optimal aspect ratio lies between 1/2 and 1 with a 30.2% sensitivity enhancement compared to the benchmark. Note that the sensitivity enhancement is specific to the given size and shape of the piezoresistor, therefore, the enhancement may be lower or higher than the obtained value depending on the practical design constraint. But in general, the sensitivity increases as the area of the piezoresistor decreases (Fig. S3 in the supplementary materials).

The sensitivity enhancement is a result of the synergy between stresses in the $x$ and $y$ directions as pointed out in the last section. To quantitatively evaluate the contribution from the two stresses, they are averaged over the piezoresistor as shown in Figs. 3(c)–(d). The average stress in the $x$ direction $\sigma_{xx,ave}$ reaches the maximum at the aspect ratio of 1/3 while $\sigma_{yy,ave}$ reaches the minimum at the aspect ratio of 1. Therefore, it is the compromise between the two leads to the optimal aspect ratio between 1/2 and 1 with a maximal sensitivity. In contrast, the benchmark model does not exhibit such aspect ratio-dependent sensitivity extrema, which demonstrates that topology optimization can generate superior designs by making full utilization of given geometric conditions.

To identify the geometric features that contribute to the high sensitivity, we measure the length $L$ of the two cantilevers and the averaged gap size measured 15, 20, and 25 μm away from the piezoresistor, $W$, as shown in Fig. 4(a). The measurements are done for a series of optimized designs shown in Fig. S4 in the supplementary materials. The measurements find that the average length $L_{ave}$ of the cantilevers increases with the aspect ratio of the piezoresistor as shown in Fig. 4(b), which is consistent with the increasing length of the piezoresistor as the aspect ratio increases. On the other hand, the average gap size $W_{ave}$ is a rapidly decreasing function of the piezoresistor aspect ratio, and it stabilizes with an average of 28.6 ± 9.6 μm when the aspect ratio is large than 1/2 as shown in Fig. 4(c).

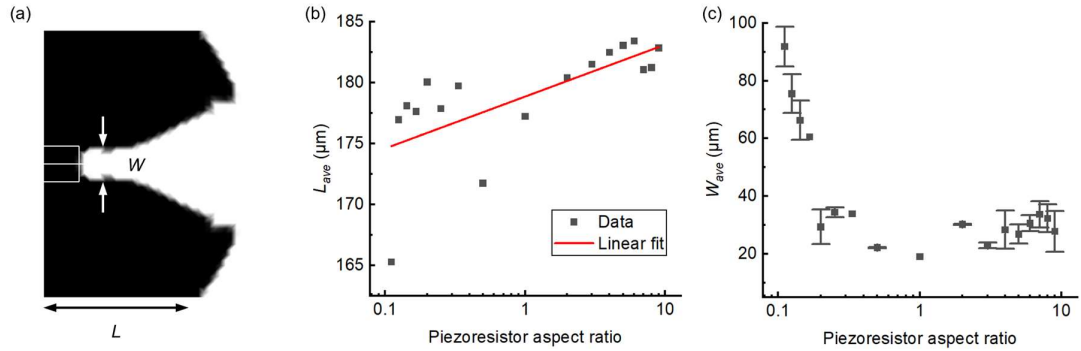

**Fig. 4.**

*3.2. A Simplified Model*

To understand the functions of the geometric features present in the optimized designs, we simplify the optimized designs into a model consisting of two cantilevers connected by a piezoresistor. This model allows us to systematically vary the geometric parameters and investigate the origin of high sensitivity in optimized designs. The aspect ratio of the piezoresistor is set to 1 for illustrative purposes. The length of the cantilevers $L$ and the gap between them $W$ are used as control parameters for the study as shown in Fig. 5(a). The distribution of stress difference for different $L$ and increasing $W$ are shown in Fig. 5(b). When there is no gap, the short cantilever has a higher sensitivity. However, as $W$ becomes nonzero, the introduction of the gap induces a highly concentrated stress difference at the intersection between two cantilevers, leading to a drastic increase in sensitivity. The sensitivity enhancement is more pronounced for the long cantilever, which is opposite to the case of rectangular cantilevers where short cantilevers have higher sensitivity. As $W$ continues to increase, the stress concentration reaches a maximum and then gradually decreases. A similar trend can also be observed in stresses in the $x$ and $y$ directions (Fig. S5 in the supplementary materials).

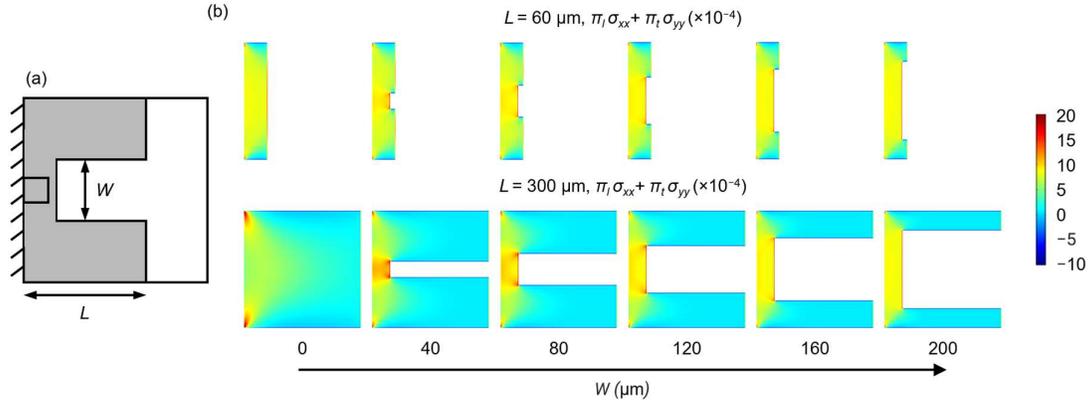

**Fig. 5.**

A comprehensive parameter sweep on $W$ and $L$ reveals that there exists an optimal gap size $W$ for each $L$ as shown in Fig. 6(a). As the cantilever lengthens, the optimal gap size shifts from 40 μm and finally stabilizes around 30 μm, which coincides with the stabilized gap sizes from topology optimization shown in Fig. 4(c). The maximal sensitivity, on the other hand, increases and saturates at a specific sensitivity as shown in Fig. 6(b). The saturated sensitivity coincides with the performance of the optimized design generated by topology optimization as marked by the horizontal line "TO". The saturation for increasing length explains the irregular shape of optimized designs shown in Fig. 3(a) since the portion longer the saturation length will not contribute to sensitivity. As a result, the flat optimization landscape leads to the random shape at the end of the cantilever, which justifies the imposed volume constraint that can regularize the optimized geometries. The average cantilever length measured in the optimized design is 177 μm, which lies at the saturated portion of the curve and is marked by the vertical line. The cantilever length in the optimized design probably results from the tradeoff between volume constraint and the diminishing return of a longer cantilever.

The coincidence of optimal gap sizes and maximal sensitivity confirms that the simplified model has captured the key features in topology optimization-generated designs. Moreover, it also serves as proof that the topology optimization has indeed generated the best design possible under the framework of double-cantilever configuration. When the two cantilevers are far away from the piezoresistor, the sensitivity decreases asymptotically to the benchmark model as marked by the dashed line denoted with "Benchmark" in Fig. 6(a), indicating that the proximity between cantilevers and the piezoresistor is essential for the high sensitivity. Note that the stresses in the $x$ and $y$ directions also exhibit the same extrema and saturation tendencies (Fig. S5 in the supplementary materials).

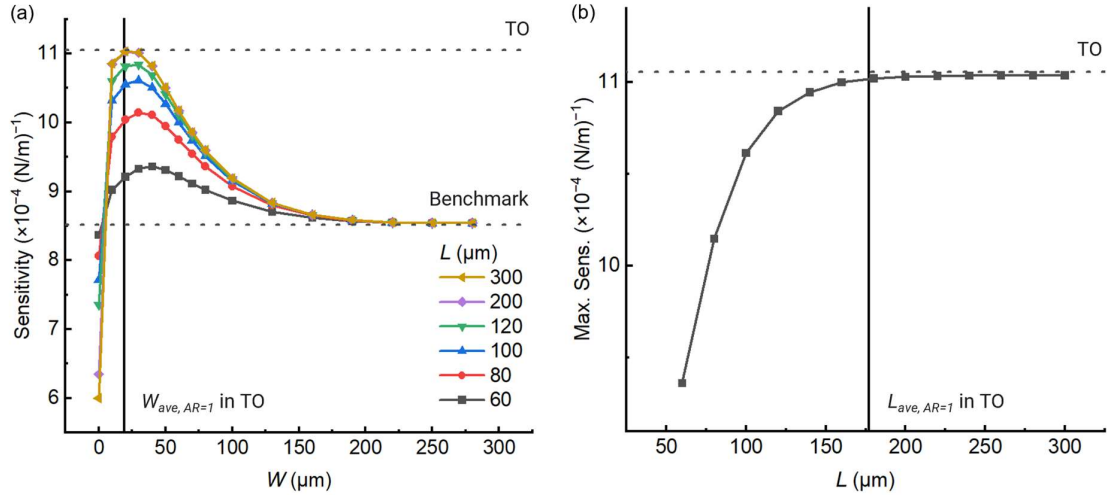

**Fig. 6.**

The parametric study on the simplified model shows a good agreement with the optimized design generated by topology optimization. The high sensitivity of the double-cantilever design stems from the gap between the two cantilevers. Essentially, the high sensitivity can be ascribed to the sudden change in bending cross-section, a design principle that has been utilized in the previously proposed design, for example, the so-called Membrane-type Surface stress Sensor (MSS).[43] One important distinction between the two designs is that MSS bends upwards under compressive surface stress, while the double-cantilever design bends downwards similar to most cantilevers. The difference in bending modes has led to their distinct optimized topologies.

Importantly for double-cantilever design, there exists an optimal gap size and cantilever length that maximize sensitivity. Both of them vary depending on the shape and size of the piezoresistor as implied by the optimized designs shown in Fig. 4(b) and Fig. S1 in the supplementary materials. Although one cannot make an assertive claim about the optimality of the double-cantilever design due to the non-convexity of the design problem, this study nonetheless provides a compelling first guess on how the optimal design may look like as the present double-cantilever design exhibits a distinct stress behavior and provides a higher sensitivity compared to conventional rectangular designs.

## 4. Conclusion

This study employed topology optimization to explore the design space of piezoresistive microcantilevers intended for use in surface stress sensing. The method allowed us to systematically generate complex geometries that tailored the stress distribution in microcantilevers with a boron-doped p-type piezoresistor oriented along the [110] direction. The results demonstrated that the optimal design depended on the shape of the piezoresistor.

We found that the highest sensitivity was achieved by using an optimal aspect ratio ranging between 0.5 and 1. Furthermore, we identified that a double-cantilever configuration was an effective design that significantly enhanced sensitivity in all the cases studied. A simplified model was developed to investigate the origin of the high sensitivity, which revealed the gap between the two cantilevers is responsible for the optimal stress distribution and high sensitivity. Moreover, the existence of optimal gap size and cantilever length indicates the distinct optimization principles between double-cantilever and conventional rectangular designs, where a smaller cantilever length is always preferred. The superior sensitivity of the double-cantilever design demonstrates the effectiveness of alternative geometries other than rectangles, providing essential knowledge for the future development of specialized designs beyond conventional cantilevers in surface stress sensing applications.

## Acknowledgments

C.Z. thanks NIMS Joint Graduate School Program, NIMS. This work was financially supported in part by the JSPS KAKENHI Grant Numbers 20K20554; 21H01971; 21K18859; 22K05324; KP19KK0141, JSPS, MEXT, Japan; and the Public/Private R&D Investment Strategic Expansion Program (PRISM), Cabinet Office, Japan.

## Figure Captions

**Fig. 1.** A schematic of the topology optimization model. (a) The 2D design domain is denoted as $\Omega$ and the obtained design is delineated by the curved boundary. The immobilization layer is above the plane, denoted as the solid curve while the silicon is below the plane, denoted as the dashed curve. (b) The optimization starts with a uniform density distribution and a piezoresistor with a fixed density of 1. As the optimization proceeds, the design gradually evolves to a discrete 0/1 optimized design.

**Fig. 2.** Benchmarking an optimized design generated by topology optimization and an optimized rectangular cantilever. (a) The density distribution of the optimized design and the rectangular cantilever. (b) The distribution of piezoresistive signal at the cantilever surface. (c,d) Stresses in the $x$ direction (c) and the $y$ direction (d).

**Fig. 3.** The optimized designs with piezoresistors of increasing aspect ratios. (a) The distribution of piezoresistive signal as a function of increasing piezoresistor aspect ratio with a fixed area $A_{piezo}$. (b) The sensitivity (averaged piezoresistive signal at the piezoresistor) of optimized designs and benchmark models as a function of piezoresistor aspect ratio. (c,d) The averaged stresses in the $x$ direction (c) and the $y$ direction (d) of optimized designs and benchmark models.

**Fig. 4.** Measuring geometrical features of optimized designs. (a) The measurement setup. (b) The average length of cantilevers in optimized designs with respect to increasing piezoresistor aspect ratios. (c) The gap sizes were measured at 15, 20, and 25 μm away from the piezoresistor. The third cantilevers at low-aspect-ratio designs are treated as gaps.

**Fig. 5.** The setup and the simulation results of the simplified model. (a) A schematic of the simplified model. (b) The distribution of piezoresistive signal under different cantilever lengths and gap sizes. When there is no gap, the short cantilever outperforms the long one. When the gap is introduced, the situation is reversed.

**Fig. 6.** Comparing the sensitivity of optimized designs and the simplified model under various parameter combinations. (a) The sensitivity of the simplified model with various plate lengths and gap sizes. The dashed horizontal lines mark the highest sensitivities of the optimized design from topology optimization (TO) and the benchmark model. The vertical line marks the measured gap size of the optimized design from TO with a square piezoresistor (aspect ratio = 1). (b) The maximal sensitivity of all plate lengths in (a). The dashed horizontal line and the vertical line mark the sensitivity and the average plate length, respectively, of the optimized design from TO.